\documentclass[%
reprint,
superscriptaddress,
nofootinbib,
 amsmath,amssymb,
prl
 aps,
]{revtex4-2}

\usepackage{graphicx}
\usepackage{dcolumn}
\usepackage{bm}

\usepackage{aas_macros}

\usepackage{orcidlink} 
\usepackage{hanging} 
\usepackage{hyperref}

\begin{document}

\preprint{APS/123-QED}

\title{Updated cosmological constraints on axion dark energy with DESI}

\author{L.~A.~Ure\~na-L\'opez\,\orcidlink{0000-0001-9752-2830}}
\affiliation{Departamento de F\'{\i}sica, DCI-Campus Le\'{o}n, Universidad de Guanajuato, Loma del Bosque 103, Le\'{o}n, Guanajuato C.~P.~37150, M\'{e}xico.}
\author{F.~Lozano-Rodríguez\,\orcidlink{0000-0001-5292-6153}}
\affiliation{Departamento de F\'{\i}sica, DCI-Campus Le\'{o}n, Universidad de Guanajuato, Loma del Bosque 103, Le\'{o}n, Guanajuato C.~P.~37150, M\'{e}xico.}
\author{J.~O.~Rom\'an-Herrera\,\orcidlink{0009-0005-5077-7007}}
\affiliation{Departamento de F\'{\i}sica, DCI-Campus Le\'{o}n, Universidad de Guanajuato, Loma del Bosque 103, Le\'{o}n, Guanajuato C.~P.~37150, M\'{e}xico.}
\author{J.~Aguilar}
\affiliation{Lawrence Berkeley National Laboratory, 1 Cyclotron Road, Berkeley, CA 94720, USA}
\author{S.~Ahlen\,\orcidlink{0000-0001-6098-7247}}
\affiliation{Physics Dept., Boston University, 590 Commonwealth Avenue, Boston, MA 02215, USA}
\author{D.~Bianchi\,\orcidlink{0000-0001-9712-0006}}
\affiliation{Dipartimento di Fisica ``Aldo Pontremoli'', Universit\`a degli Studi di Milano, Via Celoria 16, I-20133 Milano, Italy}
\affiliation{INAF-Osservatorio Astronomico di Brera, Via Brera 28, 20122 Milano, Italy}
\author{D.~Brooks}
\affiliation{Department of Physics \& Astronomy, University College London, Gower Street, London, WC1E 6BT, UK}
\author{T.~Claybaugh}
\affiliation{Lawrence Berkeley National Laboratory, 1 Cyclotron Road, Berkeley, CA 94720, USA}
\author{A.~de la Macorra\,\orcidlink{0000-0002-1769-1640}}
\affiliation{Instituto de F\'{\i}sica, Universidad Nacional Aut\'{o}noma de M\'{e}xico,  Circuito de la Investigaci\'{o}n Cient\'{\i}fica, Ciudad Universitaria, Cd. de M\'{e}xico  C.~P.~04510,  M\'{e}xico}
\author{Arjun~Dey\,\orcidlink{}}
\affiliation{NSF NOIRLab, 950 N. Cherry Ave., Tucson, AZ 85719, USA}
\author{S.~Ferraro\,\orcidlink{0000-0003-4992-7854}}
\affiliation{Lawrence Berkeley National Laboratory, 1 Cyclotron Road, Berkeley, CA 94720, USA}
\affiliation{University of California, Berkeley, 110 Sproul Hall \#5800 Berkeley, CA 94720, USA}
\author{J.~E.~Forero-Romero\,\orcidlink{0000-0002-2890-3725}}
\affiliation{Departamento de F\'isica, Universidad de los Andes, Cra. 1 No. 18A-10, Edificio Ip, CP 111711, Bogot\'a, Colombia}
\affiliation{Observatorio Astron\'omico, Universidad de los Andes, Cra. 1 No. 18A-10, Edificio H, CP 111711 Bogot\'a, Colombia}
\author{E.~Gazta\~naga}
\affiliation{Institut d'Estudis Espacials de Catalunya (IEEC), c/ Esteve Terradas 1, Edifici RDIT, Campus PMT-UPC, 08860 Castelldefels, Spain}
\affiliation{Institute of Cosmology and Gravitation, University of Portsmouth, Dennis Sciama Building, Portsmouth, PO1 3FX, UK}
\affiliation{Institute of Space Sciences, ICE-CSIC, Campus UAB, Carrer de Can Magrans s/n, 08913 Bellaterra, Barcelona, Spain}
\author{S.~Gontcho A Gontcho\,\orcidlink{0000-0003-3142-233X}}
\affiliation{Lawrence Berkeley National Laboratory, 1 Cyclotron Road, Berkeley, CA 94720, USA}
\author{G.~Gutierrez}
\affiliation{Fermi National Accelerator Laboratory, PO Box 500, Batavia, IL 60510, USA}
\author{K.~Honscheid\,\orcidlink{0000-0002-6550-2023}}
\affiliation{Center for Cosmology and AstroParticle Physics, The Ohio State University, 191 West Woodruff Avenue, Columbus, OH 43210, USA}
\affiliation{Department of Physics, The Ohio State University, 191 West Woodruff Avenue, Columbus, OH 43210, USA}
\affiliation{The Ohio State University, Columbus, 43210 OH, USA}
\author{C.~Howlett\,\orcidlink{0000-0002-1081-9410}}
\affiliation{School of Mathematics and Physics, University of Queensland, Brisbane, QLD 4072, Australia}
\author{M.~Ishak\,\orcidlink{0000-0002-6024-466X}}
\affiliation{Department of Physics, The University of Texas at Dallas, 800 W. Campbell Rd., Richardson, TX 75080, USA}
\author{R.~Kehoe}
\affiliation{Department of Physics, Southern Methodist University, 3215 Daniel Avenue, Dallas, TX 75275, USA}
\author{D.~Kirkby\,\orcidlink{0000-0002-8828-5463}}
\affiliation{Department of Physics and Astronomy, University of California, Irvine, 92697, USA}
\author{T.~Kisner\,\orcidlink{0000-0003-3510-7134}}
\affiliation{Lawrence Berkeley National Laboratory, 1 Cyclotron Road, Berkeley, CA 94720, USA}
\author{A.~Lambert}
\affiliation{Lawrence Berkeley National Laboratory, 1 Cyclotron Road, Berkeley, CA 94720, USA}
\author{M.~Landriau\,\orcidlink{0000-0003-1838-8528}}
\affiliation{Lawrence Berkeley National Laboratory, 1 Cyclotron Road, Berkeley, CA 94720, USA}
\author{L.~Le~Guillou\,\orcidlink{0000-0001-7178-8868}}
\affiliation{Sorbonne Universit\'{e}, CNRS/IN2P3, Laboratoire de Physique Nucl\'{e}aire et de Hautes Energies (LPNHE), FR-75005 Paris, France}
\author{M.~Manera\,\orcidlink{0000-0003-4962-8934}}
\affiliation{Departament de F\'{i}sica, Serra H\'{u}nter, Universitat Aut\`{o}noma de Barcelona, 08193 Bellaterra (Barcelona), Spain}
\affiliation{Institut de F\'{i}sica d’Altes Energies (IFAE), The Barcelona Institute of Science and Technology, Edifici Cn, Campus UAB, 08193, Bellaterra (Barcelona), Spain}
\author{A.~Meisner\,\orcidlink{0000-0002-1125-7384}}
\affiliation{NSF NOIRLab, 950 N. Cherry Ave., Tucson, AZ 85719, USA}
\author{R.~Miquel}
\affiliation{Instituci\'{o} Catalana de Recerca i Estudis Avan\c{c}ats, Passeig de Llu\'{\i}s Companys, 23, 08010 Barcelona, Spain}
\affiliation{Institut de F\'{i}sica d’Altes Energies (IFAE), The Barcelona Institute of Science and Technology, Edifici Cn, Campus UAB, 08193, Bellaterra (Barcelona), Spain}
\author{J.~Moustakas\,\orcidlink{0000-0002-2733-4559}}
\affiliation{Department of Physics and Astronomy, Siena College, 515 Loudon Road, Loudonville, NY 12211, USA}
\author{F.~Prada\,\orcidlink{0000-0001-7145-8674}}
\affiliation{Instituto de Astrof\'{i}sica de Andaluc\'{i}a (CSIC), Glorieta de la Astronom\'{i}a, s/n, E-18008 Granada, Spain}
\author{I.~P\'erez-R\`afols\,\orcidlink{0000-0001-6979-0125}}
\affiliation{Departament de F\'isica, EEBE, Universitat Polit\`ecnica de Catalunya, c/Eduard Maristany 10, 08930 Barcelona, Spain}
\author{G.~Rossi}
\affiliation{Department of Physics and Astronomy, Sejong University, 209 Neungdong-ro, Gwangjin-gu, Seoul 05006, Republic of Korea}
\author{E.~Sanchez\,\orcidlink{0000-0002-9646-8198}}
\affiliation{CIEMAT, Avenida Complutense 40, E-28040 Madrid, Spain}
\author{M.~Schubnell}
\affiliation{Department of Physics, University of Michigan, 450 Church Street, Ann Arbor, MI 48109, USA}
\affiliation{University of Michigan, 500 S. State Street, Ann Arbor, MI 48109, USA}
\author{J.~Silber\,\orcidlink{0000-0002-3461-0320}}
\affiliation{Lawrence Berkeley National Laboratory, 1 Cyclotron Road, Berkeley, CA 94720, USA}
\author{D.~Sprayberry}
\affiliation{NSF NOIRLab, 950 N. Cherry Ave., Tucson, AZ 85719, USA}
\author{G.~Tarl\'{e}\,\orcidlink{0000-0003-1704-0781}}
\affiliation{University of Michigan, 500 S. State Street, Ann Arbor, MI 48109, USA}
\author{B.~A.~Weaver}
\affiliation{NSF NOIRLab, 950 N. Cherry Ave., Tucson, AZ 85719, USA}
\author{H.~Zou\,\orcidlink{0000-0002-6684-3997}}
\affiliation{National Astronomical Observatories, Chinese Academy of Sciences, A20 Datun Rd., Chaoyang District, Beijing, 100012, P.R. China}

\collaboration{DESI}

\date{\today}

\begin{abstract}
We present updated constraints on the parameters of an axion dark energy model, for which we took into account the properties of its characteristic potential and its full cosmological evolution. We show that the values of the axion parameters appear sufficiently constrained by the data, including the latest DESI DR1, and are consistent with the theoretical expectations of a field mass $m_a$ in the ultralight regime $\log (m_a c^2/~\mathrm{eV}) \simeq -32.60~\pm~0.12$, and an effective energy scale $f_a$ close to the reduced Planck energy $\log (f_a/~ M_\mathrm{Pl}) \simeq -0.28~\pm~0.19$. Our results also support the idea of dynamical dark energy, although Bayesian evidence still favors the phenomenological dark energy model $w_0w_a$ over the axion dark energy, with the Bayes factor indicating moderate and weak strength of the evidence, respectively, when the models are compared to the cosmological constant $\Lambda$. However, the results suggest that axion dark energy remains a well-motivated model and its parameters may be better constrained if the upcoming DESI data show further evidence for dynamical dark energy.
\end{abstract}

\maketitle


\section{Introduction}
The Dark Energy Spectroscopic Instrument (DESI) is conducting a stage IV survey to improve constraints on dark energy models and its parameters~\cite{Snowmass2013.Levi,DESI2016a.Science,DESI2016b.Instr,DESI2022.KP1.Instr,FocalPlane.Silber.2023,Corrector.Miller.2023,FiberSystem.Poppett.2024,Spectro.Pipeline.Guy.2023,SurveyOps.Schlafly.2023}, for which it has required multiple supporting software pipelines and products~\cite{BASS.Zou.2017,LS.Overview.Dey.2019,Spectro.Pipeline.Guy.2023,SurveyOps.Schlafly.2023,SGA.Moustakas.2023}. After survey validation and early data release~\cite{DESI2023a.KP1.SV,DESI2023b.KP1.EDR}, recent DESI results suggest a dynamic nature of dark energy~\cite{DESI2024.II.KP3,DESI2024.III.KP4,DESI2024.IV.KP6,DESI:2024jis,DESI2024.VI.KP7A,DESI:2024hhd}. Under the so-called parametrization $w_0w_a$ (also known as Chevalier-Linder-Polarski)~\cite{Chevallier:2000qy,Linder:2002et,dePutter:2008wt}, the equation of state $w(a)$ of the dark energy component has the functional form $w(a) = w_0 + w_a(1-a)$, where $a$ is the scale factor of the universe, and DESI reported that $w_0 \simeq -0.7$ and $w_a \simeq -1.0$, with actual numbers slightly dependent on the considered dataset~\cite{DESI2024.VI.KP7A,DESI:2024hhd}. These results have been put under general scrutiny by the cosmological community (e.g.,~\cite{Luongo:2024fww,Gariazzo:2024sil,Carloni:2024zpl,Efstathiou:2024dvn,Park:2024jns,Dinda:2024kjf,DESI:2024aqx,DESI:2024kob,Bhattacharya:2024hep,Ramadan:2024kmn,Roy:2024kni}), in particular because of the possibility of a phantom crossing of the equation of state of dark energy (i.e., going through $w =-1$ at some point in the past). However, one cannot yet rule out dark energy models without phantom crossing; see, for instance, the discussion in~\cite{Shlivko:2024llw}, and the quintessence and models alike studied in~\cite{DESI:2024kob,Yin:2024hba,Tada:2024znt,Berghaus:2024kra,Muursepp:2024mbb,Andriot:2024jsh,Baryakhtar:2024rky,Wolf:2024stt}.

In this work, we are interested in the constraints on dark energy with an axion field after DESI DR1. The dark energy with axions was first motivated in~\cite{Frieman_1995},\footnote{We refer to the Pseudo-Nambu-Goldstone boson model, although \textit{axion} has been the standard parlance in recent years~\cite{Arvanitaki:2009fg,Hlozek:2014lca}.} and has since been studied by different groups for the past three decades~\cite{Lee:2025yvn,Abreu:2025zng,Shajib:2025tpd,Berbig:2024aee,Bhattacharya:2024kxp,Choi:2021aze,Smer_Barreto_2017,Dutta_2007,Caldwell_2005,Hall_2005,Kawasaki:2001bq,Waga_2000,Ng:2000di,Viana:1997mt,Frieman:1997xf,Coble:1996te}. Briefly, the model considers a light scalar field $\phi$ minimally coupled to gravity and endowed with a potential $V(\phi)$ of the form
\begin{equation}
  V(\phi) = m^2_a f^2_a \left[ 1+\cos\left( \phi/f_a \right) \right] \, .   \label{eq:0}
\end{equation}
We have chosen to write the potential in terms of two energy scales: $m_a$ is the axion mass and $f_a$ is the effective energy scale of the model, while also working in natural units for which $c=\hbar=k_B = 1$. 

From purely theoretical considerations~\cite{Frieman_1995}, it is expected that $m^2_a f^2_a \sim \rho_{c0}$, where $\rho_{c0}$ is the critical density of the universe, and also that $f_a \sim M_\mathrm{Pl}$, with $M_\mathrm{Pl}$ being the reduced Planck mass. In consequence, it is also expected that $m_a\sim H_0\simeq10^{-33}~\mathrm{eV}$, which then places axion dark energy in the group of ultralight axions in cosmology~\cite{Matos:2023usa,Urena-Lopez:2019kud,Hui:2016ltb}. So far, it has been possible to put valuable constraints on the amplitude of the potential~\eqref{eq:0}, which are inherited from the tight constraints on the density of dark energy: $m^2_a f^2_a \simeq 2.3 \times 10^{-11}~\mathrm{eV}^4$, but only a lower limit on the effective energy scale $f_a/M_\mathrm{Pl} > 0.67$~\cite{Smer_Barreto_2017}. 


However, the DESI results on dynamical dark energy~\cite{DESI2024.VI.KP7A} open the possibility to constrain the mass of the axion $m_a$ through the late-time evolution of the dark energy equation of state: If $m_a \gg H_0$ then the axion starts to behave as a pressureless matter component (see, for instance,~\cite{Amendola:2005ad,dePutter:2008wt,Hlozek:2014lca}), while if $m_a \ll H_0$ then its dynamics can become indistinguishable from that of a cosmological constant $\Lambda$. Both extreme cases would not be favored by the new data and the most likely output now appears to be $m_a \sim H_0$, which in turn also would allow better constraints on $f_a$. See also~\cite{Abrahamse:2007te} for an early discussion of the constraining power of stage IV experiments such as DESI on axion models.

In the remainder of this paper, we will explain the process we followed to put constraints on the axion parameters taking into account the latest DESI DR1 data, and how the results agree with the theoretical expectations of the past decades on axion models of dark energy.

\section{Background dynamics}
The equations of motion for the background evolution of a scalar field $\phi$ endowed with the potential~\eqref{eq:0}, in a homogeneous and isotropic space-time with null spatial curvature, are given by
\begin{subequations}
\label{eq:2}
  \begin{eqnarray}
    \frac{3H^2}{\kappa^2} &=& \sum_j \rho_j + \frac{1}{2} \dot{\phi}^2 + m^2_a f^2_a \left[ 1+\cos\left( \phi/f_a \right) \right] \, ,\label{eq:fried} \\
    \ddot{\phi} &=& -3 H \dot{\phi} + m_a^2 f_a \sin(\phi/f_a)  \, , \label{kge}
  \end{eqnarray}
\end{subequations}
where $\kappa^2 = 1/M^2_\mathrm{Pl}$. Also, $\rho_j$ represents the energy densities of other matter components (photons, neutrinos, baryons, cold dark matter), a dot denotes the derivative with respect to cosmic time $t$, and $H = \dot{a}/a$ is the Hubble parameter, with $a(t)$ the scale factor of the universe.

It is convenient to use a new set of polar coordinates for the axion variables, following the suggestions in~\cite{Copeland:1997et,Urena-Lopez:2015gur,Urena-Lopez:2015odd}, in the form 
\begin{subequations}
\label{eq:backvars}
\begin{gather}
\frac{\kappa \dot{\phi}}{\sqrt{6} H}  \equiv  \Omega^{1/2}_\phi \sin(\theta/2), \quad \;
 \frac{\kappa V^{1/2}}{\sqrt{3} H} \equiv \Omega^{1/2}_\phi \cos(\theta/2) \, , \\
y  \equiv \frac{2 m_a}{H} \, ,
\end{gather}
\end{subequations}
under which the Klein-Gordon equation~\eqref{kge} transforms into the following dynamical system:
\begin{subequations}
\label{eq:new4}
  \begin{eqnarray}
  \theta^\prime &=& -3 \sin \theta + \sqrt{y^2 - \alpha \Omega_\phi (1+ \cos \theta)} \, , \label{eq:new4c} \\
  y^\prime &=& \frac{3}{2}\left( 1 + w_{tot} \right) y \, , \quad  \Omega^\prime_\phi = 3 (w_{tot} + \cos\theta)
  \Omega_\phi \, , \label{eq:new4b}
\end{eqnarray}
\end{subequations}
where $\alpha = 3/\kappa^2 f_a^2$ and $\Omega_\phi = \kappa^2 \rho_\phi/3H^2 $ is the standard density parameter for the axion field. Notice that in Eq.~\eqref{eq:new4} a prime denotes derivative with respect to the number of $e$-foldings $N \equiv \ln (a/a_i)$, with $a$ the scale factor of the universe and $a_i$ its initial value, while the total equation of state is $w_{tot} = p_{tot}/\rho_{tot}$ (the ratio of the total pressure to the total density). Another quantity of interest is the equation of state of the dark energy field, $w_\phi = p_\phi/\rho_\phi$ (the ratio of the axion pressure to the axion density), which in our approach simply is $w_\phi = - \cos \theta$. This also directly shows that $-1 \leq w_\phi \leq 1$.

Equation~\eqref{eq:new4} has been presented before in Refs.~\cite{Cedeno:2017sou,LinaresCedeno:2020dte,Roy:2018nce}, being $\theta$, $y$, and $\Omega_\phi$ the new dynamical variables, with which the axion field has been studied for its useful properties as a model for dark matter and dark energy.\footnote{We do not consider density perturbations in the axion field, as they remain negligible for the whole evolution of the universe, as long as the axion field does not enter its stage of rapid oscillations. See ~\cite{Cedeno:2017sou,LinaresCedeno:2020dte,Schive:2017biq,Leong:2018opi} for a more thorough discussion of possible effects of axion fields on density perturbations.} It can be further shown that under our approach the equations of motion of the parabolic potential $V(\phi) = (1/2) m^2_a \phi^2$ are those of the dynamical system~\eqref{eq:new4} with $\alpha = 0$, see~\cite{Urena-Lopez:2015gur,Roy:2018nce}. 


\section{Bayesian analysis and parameter estimation} 
To constrain the physical parameters of the model, we use a modified version of the Boltzmann code {\scriptsize CLASS(v3.2.2)} ~\cite{Blas:2011rf} to solve Eq.~\eqref{eq:new4} for the axion field and calculate all necessary observables. 

For the numerical solution of system~\eqref{eq:new4}, we adopt the following initial conditions of the new polar variables~\cite{Roy:2018nce}: $y_i = 2m_a/H_i$ and $\theta_i = (1/5) y_i$, where the value of $m_a$ must be provided by the user. The initial value $\Omega_{\phi i}$ is adjusted using a numerical shooting routine inside the Boltzmann solver CLASS so that the desired value of the axion density parameter $\Omega_{\phi 0}$ is achieved at the present according to the Friedmann constraint~\eqref{eq:fried}. This method is very efficient at producing numerical solutions of Eq.~\eqref{eq:new4} with varied values of the cosmological parameters~\cite{Cedeno:2017sou,LinaresCedeno:2020dte,Roy:2018nce}.


We then use the software package  {\scriptsize COBAYA}~\cite{Torrado:2020dgo} for the joint process of sampling the parameters of the axion model together with the likelihoods of different observations. The priors of the different cosmological parameters are shown in Table~\ref{tab:priors}, which is the common list of parameters sampled in similar cosmological studies. Notice that some priors have been modified (see the priors in~\cite{DESI2024.VI.KP7A}) to avoid unphysical solutions of the axion model, for example, those with negative values of $\Omega_{\phi 0}$. 


\begin{table}
\caption{\label{tab:priors}List of physical parameters sampled in our study of axion models, with their corresponding priors. From top to bottom: the physical density of baryons, the physical density of dark matter, the Hubble constant, the amplitude and spectral index of primordial perturbations, the optical depth, and the two extra parameters of the axion model. Notice that the latter two are subjected to further prior constraints to narrow further the region of physical interest; see the text for more details.}
\begin{ruledtabular}
\begin{tabular}{ll}
$\omega_b$ & $\mathcal{U}[0.05:0.1]$ \\
$\omega_\mathrm{cdm}$ & $\mathcal{U}[0.1:0.2]$ \\
$H_0$ & $\mathcal{U}[50:90]$ \\
$\ln(10^{10}A_s)$ & $\mathcal{U}[1.61:3.91]$ \\
$n_s$ & $\mathcal{U}[0.8:1.2]$ \\
$\tau$ & $\mathcal{U}[0.01:0.8]$ \\
$\log(m_a c^2/\mathrm{eV})$ & $\mathcal{U}[-34:-32]$ \\
$\alpha$ & $\mathcal{U}[0:100]$ \\
\end{tabular}
\end{ruledtabular}
\end{table}

As our intention is to have a full Bayesian analysis of dark energy axion models, we chose the nested sampler {\scriptsize POLYCHORD}~\cite{Handley:2015fda,*2015MNRAS.453.4384H} (see also~\cite{Skilling:2004pqw,*Skilling:2006gxv,Ashton:2022grj,DES:2022ykc})\footnote{We used the default arguments of {\scriptsize POLYCHORD} as explained in the documentation site of {\scriptsize COBAYA}, except for $\mathrm{nlive} =10d$ and $\mathrm{nprior} = 20\mathrm{nlive}$. The latter is for a more complete exploration of the prior volume at the start of the sampling process.} to estimate the posteriors of the parameters and evidence of the models for different datasets. The latter are labeled as DESI (BAO measurements from DESI DR1), CMB (temperature and polarization data from \textit{Planck}~\cite{Planck:2019nip}, and lensing information from the combined \textit{Planck}~\cite{Carron:2022eyg}+\textit{Atacama Cosmology Telescope}~\cite{ACT:2023kun}) and SnIa (for the three separate compilations of PantheonPlus~\cite{Scolnic:2021amr,*Brout:2022vxf}, Union3~\cite{Rubin:2023ovl} and DESY5~\cite{DES:2024tys}). With DESI+CMB the common dataset, we will henceforth use the SnIa compilations to distinguish the three separate datasets: +PantheonPlus (DESI + CMB + PantheonPlus), +Union3 (DESI + CMB + Union3), and +DESY5 (DESI + CMB + DESY5).

In turn, a note is made regarding the prior of our model. Some combinations of the pair $(m_a,\alpha)$ in the prior plane $[-34:-32] \times [0:100]$ of Table~\ref{tab:priors} can also lead to solutions outside the region of interest for our purposes, for example, solutions in which the axion model has a mass large enough, e.g., $m_a \gg H_0$, that it is starting to behave as dark matter at recent times. It is desirable to avoid such combinations of values $(m_a,\alpha)$ in the prior volume, since an unrepresentative prior (in our case, one with large regions devoid of dark-energy solutions) can result in inefficient sampling of the parameters and estimation of the evidence (see also~\cite{Ashton:2022grj,chen2022bayesianposteriorrepartitioningnested} for a broader discussion of priors and nested samplers).

To overcome this difficulty, it is necessary to consider a more informative prior in the variables $(m_a,\alpha)$, and for this, we take inspiration from the recent DESI results~\cite{DESI2024.VI.KP7A}. The constraints on the parameters $(w_0,w_a)$ can also be seen as constraints on the present values of the equation of state of dark energy $w$ and its derivative $w^\prime = dw/d(\ln a)$, namely, $w_0$ and $w^\prime_0$, respectively. For the $w_0 w_a$ model, we find that the parameter $w_0$ is already the present value of the equation of state, and also that $w^\prime_0 = -w_a$. Thus, we can make the interpretation that the DESI results on $(w_0,w_a)$ mean a preference of cosmological data for the region $[-1:0] \times [0:3]$ in the plane $(w_0,w^\prime_0)$.

Our strategy is then to select the values of $(m_a,\alpha)$ for which we obtain the points $(w_{\phi 0}, w^\prime_{\phi 0})$ in the extended plane $[-1:0] \times [0:8]$.~\footnote{Recall that $-1 \leq w_\phi \leq 1$, then we are choosing the values $-1 \leq w_{\phi 0} \leq 0$ that most likely correspond to an accelerated expansion of the universe, while $w^\prime_{\phi 0} \geq 0$ means that the axion field is in the thawing regime~\cite{Caldwell_2005,dePutter:2008wt}.} This can be achieved by including a new likelihood function $\mathcal{L}(w_{\phi 0},w^\prime_{\phi 0})$ such that $\mathcal{L}(w_{\phi 0},w^\prime_{\phi 0}) =1$ if the derived values of $(w_{\phi 0},w^\prime_{\phi 0})$ are in the region ($[-1:0] \times [0:8]$), and $\mathcal{L}(w_{\phi 0},w^\prime_{\phi 0}) =0$ otherwise. This change in the likelihood of our model does not imply larger times for running {\scriptsize POLYCHORD}, but only some extra time to set the true prior for the values of $(m_a,\alpha)$ and initialize the sampler.

In turn, another prior constraint is applied. For a fixed value of $m_a$, a higher value of $\alpha$ means that the initial value $\phi_i$ is closer to the top of the cosine potential, that is, $\phi_i/f_a \to 0$ as $\alpha \to \infty$. In this limit, the field density $\rho_\phi$ remains constant up to the present time for any value of $m_a$. However, the numerical solution becomes more difficult for larger values of $\alpha$, and there is a higher risk of failures in the Boltzmann solver {\scriptsize CLASS}. To avoid that, we also impose the prior condition $\phi_i/f_a > 0.01 \simeq 0.6^\circ$ (see also the extended discussion in~\cite{LinaresCedeno:2020dte}).

\section{Results and discussion}
\begin{figure*}[tp!]
\includegraphics[width=0.48\textwidth]{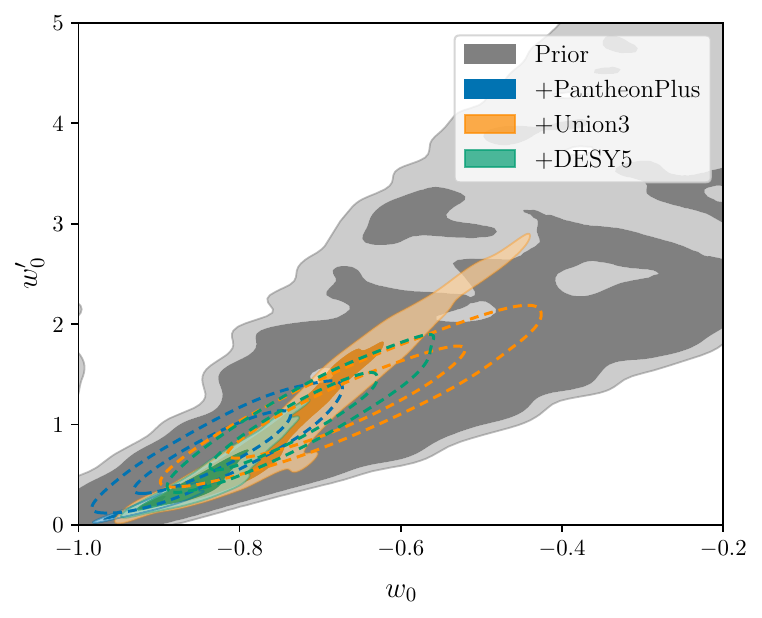}
\includegraphics[width=0.48\textwidth]{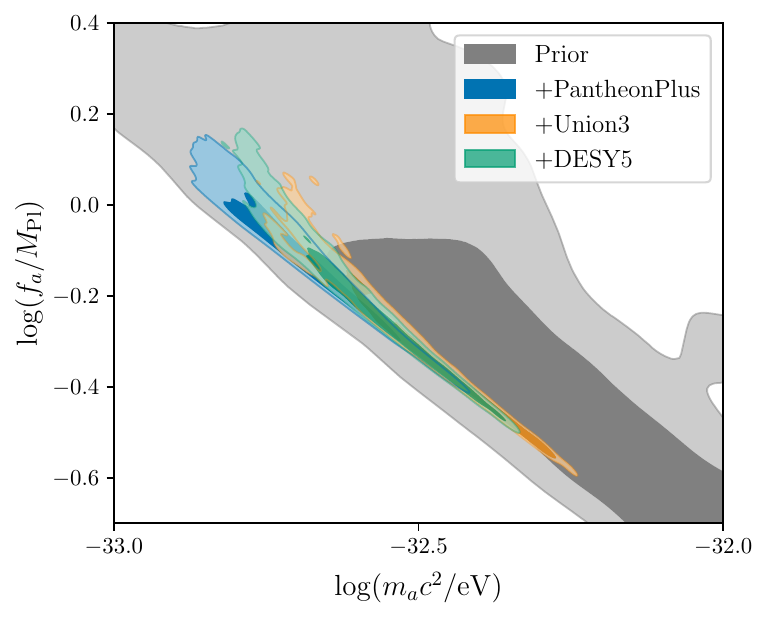}
\caption{Left: 95\% credible regions of the present values of the equation of state $w_0$ and its logarithmic derivative $w^\prime_0$. The solid contours correspond to the axion field, while the line contours correspond to the $w_0w_a$ model. The colors indicate the different datasets described in the text. Right:  95\% CL of the axion parameters $m_a$ and $f_a$, both on logarithmic scale and with the same colors and datasets. Also shown on the background in both panels are the prior distributions of the corresponding parameters.}
\label{fig:posteriors}
\end{figure*}

Our main results are synthesized in the posterior distributions of the parameters shown in Fig.~\ref{fig:posteriors}, and in the estimated values of the parameters at $95\%$ CL listed in Table~\ref{tab:results} (using {\scriptsize GETDIST}~\cite{lewis2019getdistpythonpackageanalysing}). Taking advantage of nested sampling, the posterior distributions are plotted on top of the constrained prior ones, which were also estimated by {\scriptsize POLYCHORD} at the beginning of the sampling. (Constrained priors refer to the prior region where the likelihood exceeds a given threshold~\cite{Ashton:2022grj}.)

First, we show on the left panel of Fig.~\ref{fig:posteriors} the credible regions of the derived parameters $(w_0,w^\prime_0)$ for the axion model, together with those corresponding to the $w_0w_a$ model for the same datasets and priors of common parameters. Interestingly enough, the credible levels of $(w_0,w^\prime_0)$ in Table~\ref{tab:results} also show a preference for values that are away from those of $\Lambda$, following a trend similar to those of the model $w_0w_a$. Likewise, the ratio $-w^\prime_0/(1+w_0) \simeq 2.83$ suggests a preference for values closer to the so-called \textit{mirage} regime ($\simeq -3.66$)~\cite{Linder:2007ka} than to the so-called \textit{thawing} regime ($\simeq -1.58$)~\cite{Linder:2015zxa} (see also~\cite{DESI:2024kob}).

\begin{table*}
\caption{\label{tab:results}List of physical parameters and their constraints at $95\%$ CL using different datasets. From left to right, the columns show the corresponding datasets of the sampling (as described in the text), the fractional density of matter $\Omega_m$, the present value of the Hubble parameter $H_0$ (in units of $\mathrm{km} \, \mathrm{s}^{-1} \mathrm{Mpc}^{-1}$), the axion mass $m_a c^2$, the effective energy scale $f_a$, the present values of the axion equation of state $w_0$, its logarithmic derivative $w^\prime_0$, and their ratio. Also shown for comparison are the values of the corresponding quantities in the models $w_0w_a$ and $\Lambda$.}
\begin{ruledtabular}
\begin{tabular}{lccccccc}
Dataset & $\Omega_m$ & $H_0$ & $\log (m_a c^2/\mathrm{eV})$ & $\log(f_a/M_\mathrm{Pl})$ & $w_0$ & $w^\prime_0$ & $-w^\prime_0/(1+w_0)$ \\
\hline
+PantheonPlus & $0.314^{+0.011}_{-0.010}$ & $67.0^{+0.9}_{-1.0}$ & $-32.69^{+0.18}_{-0.18}$ & $-0.11^{+0.24}_{-0.22}$ & $-0.92^{+0.06}_{-0.05}$ & $0.19^{+0.18}_{-0.15}$ & $-2.2^{+0.6}_{-0.7}$ \\
$[w_0w_a]$  & $0.308^{+0.013}_{-0.013}$ & $68.1^{+1.5}_{-1.4}$ & $\cdots$ & $\cdots$ & $-0.83^{+0.13}_{-0.12}$ & $0.74^{+0.54}_{-0.52}$ & $-4.5^{+6.7}_{-6.6}$ \\
$[\Lambda]$  & $0.307^{+0.010}_{-0.010}$ & $67.9^{+0.7}_{-0.7}$ & $\cdots$ & $\cdots$ & $-1$ & $0$ & $\cdots$ \\
\hline 
+Union3 & $0.330^{+0.018}_{-0.016}$ & $65.3^{+1.5}_{-1.8}$ & $-32.48^{+0.20}_{-0.24}$  & $-0.33^{+0.31}_{-0.23}$ & $ -0.75^{+0.22}_{-0.18}$ & $0.97^{+1.50}_{-0.87}$ & $-3.5^{+1.5}_{-1.7}$ \\
$[w_0w_a]$  & $0.322^{+0.018}_{-0.017}$ & $66.7^{+1.8}_{-1.8}$ & $\cdots$ & $\cdots$ & $-0.66^{+0.19}_{-0.18}$ & $1.23^{+0.74}_{-0.70}$ & $-3.7^{+0.9}_{-1.0}$ \\
$[\Lambda]$  & $0.307^{+0.010}_{-0.010}$ & $67.9^{+0.7}_{-0.7}$ & $\cdots$ & $\cdots$ & $-1$ & $0$ & $\cdots$ \\
\hline
+DESY5 & $0.322^{+0.014}_{-0.013}$ & $66.2^{+1.1}_{-1.2}$ & $-32.58^{+0.20}_{-0.21}$ & $-0.22^{+0.31}_{-0.26}$ &  $-0.84^{+0.10}_{-0.08}$ & $0.45^{+0.48}_{-0.33}$ & $-2.8^{+1.0}_{-1.2}$ \\
$[w_0w_a]$  & $0.316^{+0.013}_{-0.013}$ & $67.3^{+1.3}_{-1.3}$ & $\cdots$ & $\cdots$ & $-0.73^{+0.14}_{-0.13}$ & $1.05^{+0.63}_{-0.60}$ & $-3.8^{+1.0}_{-1.0}$ \\
$[\Lambda]$  & $0.310^{+0.010}_{-0.010}$ & $67.7^{+0.7}_{-0.7}$ & $\cdots$ & $\cdots$ & $-1$ & $0$ & $\cdots$
\end{tabular}
\end{ruledtabular}
\end{table*}



Second, on the right panel we show the contours for the physical parameters of the axion model, $(m_a,f_a)$. This is the main result of this work, as it is the first time that closed contours are found for these physical parameters. As expected, the constrained values of the mass of the axion lie around $\log (m_a c^2/~\mathrm{eV}) \simeq -32.60 \pm 0.12$, while for the effective energy scale of the axion, we find $\log(f_a/~M_\mathrm{Pl}) \simeq -0.28 \pm 0.19$. This is the first time that the effective energy scale of an axion, with a value close to, but smaller than, the Planck energy, has been constrained. Similar results have been obtained in recent studies with approximate methods to solve the axion equations of motion~\cite{Bhattacharya:2024hep,Berbig:2024aee}.


We now use Bayesian model selection to have an assessment of axion dark energy compared to $\Lambda$ and $w_0w_a$. With {\scriptsize ANESTHETIC}~\cite{anesthetic}, we calculated the following three quantities for each model: evidence $\mathcal{Z}$, Kullback-Leibler divergence $\mathcal{D}_\mathrm{KL}$ (which quantifies the overall compression from prior to posterior distribution, see also Fig.~\ref{fig:posteriors}), and the posterior average of the log likelihood $\langle \ln \mathcal{L} \rangle_\mathcal{P}$ (which gives us an estimate of the goodness of fit of the model to the data). They are related by Bayes' theorem in the form $\ln \mathcal{Z} = \langle \ln \mathcal{L} \rangle_\mathcal{P} - \mathcal{D}_\mathrm{KL}$~\cite{Hergt:2021qlh,Trotta:2008qt}, which tells us that $\mathcal{D}_\mathrm{KL}$ acts as an Occam penalty on the model: in log units, the evidence is given by the goodness of fit minus the compression from prior to posterior. The latter then explicitly shows the dependence of evidence $\mathcal{Z}$ on the choice of priors for the parameters in the model.

The Bayes factor for each model $\mathcal{M}$ is given by formula $\ln B_{\mathcal{M}\Lambda} = \ln \mathcal{Z}_\mathcal{M} - \langle \ln \mathcal{Z}_\Lambda \rangle$, where $\langle \ln \mathcal{Z}_\Lambda \rangle$ is the mean value of the evidence of the model with $\Lambda$ (again with the same priors for the common parameters). The Bayes factor will be interpreted in terms of Jeffrey's scale in~\cite{Trotta:2008qt} for the strength of the evidence: $|\ln B_{\mathcal{M}\Lambda}| < 1$ is inconclusive; $|\ln B_{\mathcal{M}\Lambda}| = 1$ is weak; $|\ln B_{\mathcal{M}\Lambda}| = 2.5$ is moderate; and $|\ln B_{\mathcal{M}\Lambda}| = 5$ is strong. Also, if $\ln B_{\mathcal{M}\Lambda} > 0$ ($\ln B_{\mathcal{M}\Lambda} < 0$) then the evidence is in favor of (against) model $\mathcal{M}$ compared to $\Lambda$.

All of the above quantities are shown in Table~\ref{tab:bayes}, where, similarly to the Bayes factors, $\langle \ln \mathcal{L} \rangle_\mathcal{P}$ and $\mathcal{D}_\mathrm{KL}$ are also given in terms of their difference with respect to the corresponding mean quantity for $\Lambda$.

\begin{table}
\caption{\label{tab:bayes}The Bayes factor of the axion ($B_{a \Lambda}$) and $w_0w_a$ ($B_{w_0w_a \Lambda}$) models with respect to $\Lambda$, calculated as explained in the text for each of the same datasets as in Fig.~\ref{fig:posteriors} and Table~\ref{tab:results}. Also shown are $\langle \ln \mathcal{L} \rangle_\mathcal{P}$ and $\mathcal{D}_\mathrm{KL}$, in terms of their difference with respect to the same quantities for $\Lambda$. Note that, due to the use of different priors in Table~\ref{tab:priors}, the values of the Bayes factor $\ln B_{w_0w_a \Lambda}$ differ slightly from those reported in~\cite{DESI2024.VI.KP7A}.}
\begin{ruledtabular}
\begin{tabular}{lccc}
 & +PantheonPlus & +Union3 & +DESY5 \\
\hline
Axion \\
$\ln B_{a \Lambda}$ & $-4.2 \pm 0.5$ & $0.9 \pm 0.4$ & $0.6 \pm 0.4$ \\
$\Delta \langle \ln \mathcal{L} \rangle_\mathcal{P}$ & $0.8 \pm 0.2$ & $1.0 \pm 0.2$ & $3.1 \pm 0.2$ \\
$\Delta \mathcal{D}_\mathrm{KL}$ & $5.0 \pm 0.5$ & $0.1 \pm 0.4$ & $2.5 \pm 0.4$ \\
\hline
$w_0 w_a$ \\
$\ln B_{w_0w_a \Lambda}$ & $-2.5 \pm 0.4$ & $2.5 \pm 0.4$ & $3.0 \pm 0.6$ \\
$\Delta \langle \ln \mathcal{L} \rangle_\mathcal{P}$ & $3.1 \pm 0.2$ & $6.3 \pm 0.2$ & $7.6 \pm 0.2$ \\
$\Delta \mathcal{D}_\mathrm{KL}$ & $5.6 \pm 0.4$ & $3.8 \pm 0.4$ & $4.6 \pm 0.4$
\end{tabular}
\end{ruledtabular}
\end{table}

In general terms, we find that the data do not show a preference of the axion model over $\Lambda$, with evidence at most in the weak regime, according to Jeffrey's scale, while there is still a moderate preference for the data of the $w_0w_a$ parametrization over the axion and $\Lambda$ models. Notice that both the axion and $w_0 w_a$ models have better goodness of fit to each of the datasets than $\Lambda$, as expected from models with more free parameters. But they also suffer from larger Occam penalties, mainly because the posterior is sufficiently informative (the free parameters appear to be well constrained) and has an uncertainty smaller than that of the prior~\cite{Trotta:2008qt}. However, it is the $w_0w_a$ model that more easily overcomes the Occam penalty and gets the best evidence.

We have presented updated constraints on the axion model of dark energy, and show for the first time that cosmological data provide separate constraints on the free parameters $(m_a,f_a)$. It is also confirmed that the values of the latter are consistent with theoretical expectations of an axion mass in the ultralight regime and an effective energy scale close to the reduced Planck energy. Our results also support the idea of dynamical dark energy under the axion hypothesis, with slightly better fit to the data than $\Lambda$ (see $\langle \ln \mathcal{L} \rangle_\mathcal{P}$ in Table~\ref{tab:bayes}), although Bayesian evidence still favors the phenomenological model $w_0w_a$.\footnote{Reference \cite{Wolf:2024eph} considered a field model with $V(\phi) = V_0 \pm (1/2) m^2_a \phi^2$, and concluded that the cosmological data favor the hilltop option with $-m^2$. The latter case is a second-order approximation of the potential~\eqref{eq:0} around its top point. Reference \cite{Wolf:2024eph} also notes that evidence favors $\Lambda$ for the combination +PantheonPlus and that the field model cannot overcome the $w_0 w_a$ one. This coincides with our result in Table~\ref{tab:bayes}, although, as we report there, the evidence for the field model improves with other SNIa data.} 

Nevertheless, axion dark energy remains a compelling model,\footnote{ Axion dark energy provides a natural explanation~\cite{Visinelli:2017imh,Fujita:2020ecn,Lee:2025yvn,Ballardini:2025apf} of the isotropic cosmic birefringence angle $\beta = 0.34 \pm 0.09 \, \mathrm{deg}$ reported in~\cite{Diego-Palazuelos:2022dsq,Eskilt:2022wav,Eskilt:2022cff,Cosmoglobe:2023pgf,Naokawa:2025shr}. The angle $\beta$ is related to the field cosmic excursion $\Delta \phi$ by $\beta = (g f_a/2) (\Delta \phi/f_a)$, where $g$ is the (parity-violating) axion-photon coupling constant (e.g.~\cite{Fujita:2020ecn,Choi:2021aze,Namikawa:2025sft,Eberhardt:2025caq}). Our results suggest that $\Delta\phi/f_a \simeq 0.3$ (see also~\cite{DESI:2025fii}), and then $g f_a = \mathcal{O}(0.1)$.} and it will be interesting to repeat this study with the forthcoming DESI data (e.g.~\cite{DESI:2025fii}) and see whether the parameters of the model are better constrained and the Bayesian evidence improves compared to other scenarios than those reported here.

\begin{acknowledgments}
F. L.-R. and J. O. R.-H. acknowledge SECIHTI for support from doctoral fellowships. L. A. U.-L. acknowledges partial support from the Programa para el Desarrollo Profesional Docente; Direcci\'on de Apoyo a la Investigaci\'on y al Posgrado, Universidad de Guanajuato, under Grants No. 046/2024 and No.  334/2025; and SECIHTI Mexico under Grants No. 304001 and No. CBF-2025-G-1327. L. A. U.-L. is grateful to Nandan Roy and Burin Gumjudpai for their kind hospitality and useful discussions in a visit to the Centre for Theoretical Physics \& Natural Philosophy, Mahidol University, Thailand, where part of this work was developed.

The authors are honored to be permitted to conduct scientific research on Iolkam Du’ag (Kitt Peak), a mountain with particular significance to the Tohono O’odham Nation. This material is based upon work supported by the U.S. Department of Energy (DOE), Office of Science, Office of High-Energy Physics, under Contract No.\ DE--AC02--05CH11231, and by the National Energy Research Scientific Computing Center, a DOE Office of Science User Facility under the same contract. Additional support for DESI was provided by the U.S. National Science Foundation (NSF), Division of Astronomical Sciences, under Contract No.\ AST-0950945 to the NSF's National Optical-Infrared Astronomy Research Laboratory; the Science and Technology Facilities Council of the United Kingdom; the Gordon and Betty Moore Foundation; the Heising-Simons Foundation; the French Alternative Energies and Atomic Energy Commission (CEA); the National Council of Humanities, Science and Technology of Mexico (CONAHCYT); the Ministry of Science, Innovation and Universities of Spain (MICIU/AEI/10.13039/501100011033), and by the DESI Member Institutions: \url{https://www.desi.lbl.gov/collaborating-institutions}.

Any opinions, findings, and conclusions or recommendations expressed in this material are those of the author(s) and do not necessarily reflect the views of the U. S. National Science Foundation, the U. S. Department of Energy, or any of the listed funding agencies.
\end{acknowledgments}

\vspace{-0.4cm}\section*{Data availability}
\vspace{-0.4cm}
 The data that support the findings of this article are openly available in~\cite{zenodo}.


\bibliography{axionbib}

\end{document}